\documentstyle[11pt,appb,epsf]{article}
\newcommand{\mylims}{\stackrel{\scriptscriptstyle N\rightarrow\infty}
{\longrightarrow}}
\begin{document} 
\title{TOWARDS NEW UNDERSTANDING OF NUCLEAR ROTATION
\thanks{Presented at the XXXIII Zakopane School of Physics,
Zakopane, Poland, September 1 -- 9, 1998.} }
\author{\underline{P.~Rozmej}$^{\rm a}$ and R.~Arvieu$^{\rm b}$
\address{
$^{\rm a}$Theoretical Physics Department, University MCS, 20-031 Lublin, Poland\\
$^{\rm b}$Institut des Sciences Nucl\'eaires, F 38026 Grenoble-Cedex, France\\}
}

\maketitle

\begin{abstract}
Properties of time evolution of wave packets built up from
rotator eigenstates are discussed. The mechanism of perfect cloning
of the initial wave packet for {\em circular states} at fractional
revival times is explained. The smooth transition from
{\em circular} to {\em linear} through intermediate {\em elliptic states}   
is described. Examples of time evolution of a nuclear wave packet created in
Coulomb excitation mechanism are presented.     
\end{abstract}

\section{Introduction}

The advent of experimental techniques (mainly short, tunable laser pulses)
resulted in rapidly growing interest in time evolution of wave packets (WP)
in many fields of physics and chemistry. Nine years ago Averbukh and 
Perelman {aver} discovered and explained a universal scenario
of this time evolution applicable to a wide class of quantum systems.
The general requirements for their scenario are very weak: 
{\em i)} the initial WP have to be a superposition of bound states
of system's hamiltonian, {\em ii)} the weights of the superposition should be 
strongly peaked around the mean value ($\psi(t=0)=\sum_n\, c_n \phi_n$ where
$H\phi_n=E_n\phi_n$).
In such conditions, as we already explained during the previous conference
\cite{zak96}, the evolution is governed basically by two time scales, 
$T_{cl}=2\pi\hbar/|E'_n|_{n=\bar{n}}$ -- the period of motion of the
corresponding classical system with $E=E_{\bar{n}} $, and 
$T_{rev}=2\pi\hbar/|E''_n|_{n=\bar{n}}$ -- revival time, after which the
WP reassembles to a shape reminding the initial one. In general the 
longer time scales also exist and usually $T_{cl}<<T_{rev}<<\ldots $.
However, there exist systems where $E_n$ depends quadratically on the quantum
number $n$ (infinite square well, ideal rotator). In such cases the 
quantum evolution is exactly periodic with $T=T_{rev}$.
The general scenario of \cite{aver} exhibits in these cases some particularly
interesting features.
\section{Time evolution of WP in quantum rotator, clones and mutants}

The time evolution of the WP is given by the formula:
\begin{equation} \label{ev1}
\Psi(t) = \sum_n \,c_n\,\phi_n\, \exp(-iE_n t/\hbar)\;.
\end{equation}
Let us expand energies in Taylor series $E_n=E_{\bar{n}}+
E'_{\bar{n}}\,(n-\bar{n})+E''_{\bar{n}}\,(n-\bar{n})^2+\ldots$ 
If $E_n$ is only quadratic function of quantum number $n$, the series 
terminates at $E''_{\bar{n}}$. This is the case for quantum rotator, 
$H=(\hbar^2/2J)\,I^2$, with $E_I=(\hbar^2/2J)\,I(I+1)$.
Introducing time scales mentioned above one can rewrite (\ref{ev1})
in the form $(k=(I-\bar{I}))$:
\begin{equation} \label{ev2}
\Psi(t) = \sum_I \,c_I\,\phi_I\, \exp[-2\pi i(kt/T_{cl}+k^2 t/T_{rev})] \;\;.
\end{equation}
For all times of the form $t=(m/n)T_{rev}$ (where $m$ and $n$ are
mutually prime numbers) it is possible to use Gauss sum and rewrite
(\ref{ev2}) in the form:
\begin{equation} \label{evcl}
\Psi(t=\frac{m}{n}T_{rev}) = \sum_{s=0}^{l-1} \, a_s\, \Psi_{cl}^s \;\;,
\end{equation}
where $l=n/2$ for $n$--multiple of 4, $l=n$ otherwise, $\Psi_{cl}^s$
is either identical ({\em clone}) or similar ({\em mutant}) to the initial WP, 
depending on topology of the motion.

\begin{figure}[bh] \vspace{-132mm} \hspace{-6mm}
\epsfxsize=116mm \epsfbox{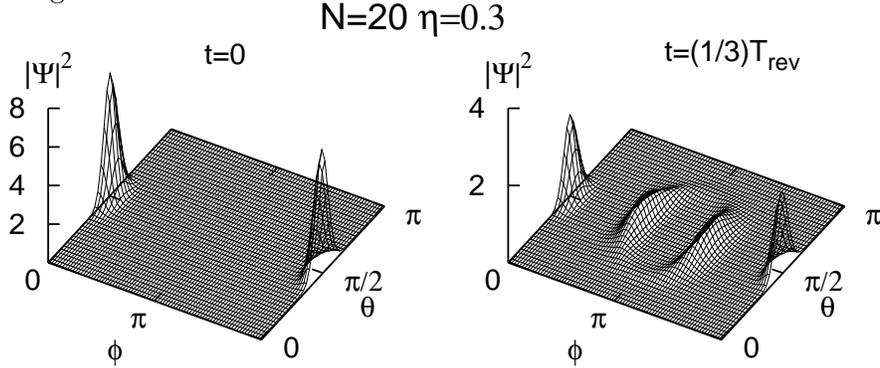}
\vspace{5mm}
\caption{Initial WP (left) and evolved one at $t=(1/6)T_{rev} \mbox{~or~}
(1/3)T_{rev}$ (right). One clone and two mutants (crescent shaped)
are clearly visible in the right figure. } \label{clonmut}
\end{figure}

The initial WP can be created as a coherent state (CS) of angular momentum
\cite{roz,arv},
fulfilling during evolution the minimum uncertainty condition:
$\Delta L_x^2\, \Delta L_y^2 = \frac{1}{4}\langle L_z \rangle^2$.
One of the best possible choice of such CS can be written in an exponential form
depending on 2 real parameters $N$ and $\eta$:
\begin{equation} \label{psineta}
\Psi_{N,\eta}(\theta,\phi) = \sqrt{\frac{N}{2\pi\sinh(2N)}}\,
\mbox{e}^{N\sin\theta(cos\phi+i\eta\sin\phi)}  \;\;,
\end{equation}
with $ \langle L_z \rangle\mylims \, \eta\,(N-\frac{1}{2})$ and 
$\eta = {\langle L_z \rangle}/(2\Delta L_y^2) = 
\pm \sqrt{\Delta L_x^2/\Delta L_y^2} $. 
There are 2 special cases, particularly
interesting: {\em i)} $\eta=1$, corresponding to {\em circular states}, and
{\em ii)} $\eta=0$, so called {\em linear states}.
In the former case the expansion of the initial WP in spherical harmonics
contains only functions with maximal $M=I$,
in the latter only those with $M=0$.
In general case, $\eta\neq 0$, additional sumation over $M$ is necessary
and the motion is called {\em elliptic}:
\begin{equation} \label{ellip}
|CS,\eta,t=0\rangle = \sum_{IM} \, b_{IM}(N,\eta)\, Y^I_M(\theta,\phi)  \;\;.
\end{equation}
In all cases the coefficients $b_{IM}$ are given analytically.
For $\eta=1$ all fractional revivals (\ref{evcl}) are copies of the initial WP
(clones). Their number is equal $q=n/2$ for even $n$ or $q=n$ for odd $n$.
For $\eta\neq 1$ fractional revivals have in general different shapes 
(from crescents on a sphere to rings at $\eta=0$) than 
the initial WP (we call them mutants). However if the position of fractional
revival coincides with that of the initial WP the clone is always built.  
The example of WP shapes for the {\em elliptic state} with $N=20, \eta=0.3$
is shown in Fig. \ref{clonmut}.
\section{Nuclear rotation}

\begin{figure}[t] \vspace{-80mm}  \hspace{2mm}
\epsfxsize=100mm \epsfbox{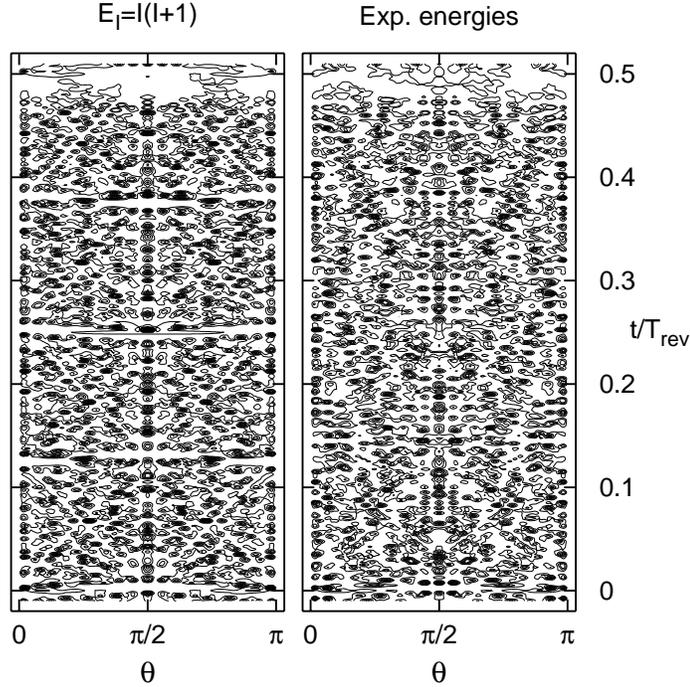}
\vspace{-25mm}
\caption{Time evolution of nulear rotational wave packet obtained in CE of
$^{238}$U presented in `quantum carpet' representation. Contours of
$2\pi\sin\theta |\Psi|^2$ are plotted.} \label{carp}
\end{figure}

It is well known \cite{broglia,fonda} that during Coulomb excitation (CE)
a deformed nucleus is excited to a coherent mixture of rotational states.
This superposition is also peaked around a mean value of angular momentum,
so one can expect similar features as predicted by scenario of Averbukh and
Perelman. The most clear case is CE with backscattering as in this case 
excited WP has cylindrical symmetry (only $Y^I_0$ components, as {\em linear}
CS). The partial waves and fractional revivals have then
topology of rings on a sphere. Therefore for presentation of shapes only
one angular variable ($\theta$) is sufficient. In Fig. \ref{carp}
we present the time evolution of WP obtained by CE of $^{238}$U bombarded by
$^{40}$Ar at E=170 MeV. The amplitudes of excitation of given $I$ angular momentum
eigenstates have been calculated within semiclassical theory of CE \cite{broglia}.
The left part shows the `ideal case', i.e. when $E_I$'s follow perfect rotor
dependence $I(I+1)$, the right corresponds to time evolution (\ref{ev1}) 
with energies taken from experiment.  
Although the 'carpet' for experimental energies isn't as regular as
that for `ideal' ones, still strong revivals of WP occur. The absolute
time scales for nuclear rotation are very short ($T_{rev}\sim 10^{-19}$s, 
$T_{cl}\sim 10^{-20}$s) and are still beyond time resolution of present
experimental techniques. For more details see \cite{roz,arv}.

\end{document}